\documentclass[usenatbib]{mn2e}

\usepackage{graphicx,url}
\usepackage{amssymb,amsmath}
\usepackage[pass,a4paper]{geometry}

\title[Glitches have a minimum size]{Neutron star glitches have a substantial minimum size}
\author[C.~M. Espinoza et al.]{C.M.~Espinoza,$^{1, 2}$\thanks{E-mail: cespinoz@astro.puc.cl}
D.~Antonopoulou,$^{3}$
B.W.~Stappers,$^{1}$
A.~Watts$^{3}$ and 
A.G.~Lyne$^{1}$ \\
$^{1}$Jodrell Bank Centre for Astrophysics, School of Physics and Astronomy,
The University of Manchester, Manchester M13 9PL, UK. \\
$^{2}$Instituto de Astrof\'isica, Facultad de F\'isica, Pontificia Universidad Cat\'olica de Chile, Casilla 306, Santiago 22, Chile. \\
$^{3}$Astronomical Institute Anton Pannekoek, University of Amsterdam, Postbus 94249, 1090GE Amsterdam, The Netherlands. \\
}

\date{\today}
\pagerange{\pageref{firstpage}--\pageref{lastpage}} \pubyear{2012}

\begin{document}

\label{firstpage}
\maketitle
\begin{abstract}
Glitches are sudden spin-up events that punctuate the steady spin down of pulsars and are thought to be due to the presence of a superfluid component within neutron stars. 
The precise glitch mechanism and its trigger, however, remain unknown.  
The size of glitches is a key diagnostic for models of the underlying physics. 
While the largest glitches have long been taken into account by theoretical models, it has always been assumed that the minimum size lay below the detectability limit of the measurements.   
In this paper we define general glitch detectability limits and use them on 29 years of daily observations of the Crab pulsar, carried out at Jodrell Bank Observatory.  
We find that all glitches lie well above the detectability limits and by using an automated method to search for small events we are able to uncover the full glitch size distribution, with no biases. 
Contrary to the prediction of most models, the distribution presents a rapid decrease of the number of glitches below $\sim0.05\,\mu$Hz. 
This substantial minimum size indicates that a glitch must involve the motion of at least several billion superfluid vortices and provides an extra observable which can greatly help the identification of the trigger mechanism.
Our study also shows that glitches are clearly separated from all the other rotation irregularities.
This supports the idea that the origin of glitches is different to that of timing noise, which comprises the unmodelled random fluctuations in the rotation rates of pulsars.

\end{abstract}


\section{Introduction}
Neutron stars are the highly-magnetised and rapidly-rotating remnants of the collapse of the cores of once more-massive stars. 
Having masses of approximately $1.4\,\rm{M}_\odot$ and radii of about $12$\,km, the high densities of neutron stars indicate a structure of a crystalline-like crust and a superfluid interior \citep{bpp69,NS1}. 
Their large and steady moments of inertia mean that they have extremely stable rotational frequencies, which slowly decrease as energy is lost through electromagnetic radiation and acceleration of particles in their magnetospheres.  
However, this regular spin-down is occasionally interrupted by sudden spin-up events, known as glitches \citep{rm69,elsk11}.

The exact mechanism responsible for glitches is not fully understood but it is thought to involve a sudden transfer of angular momentum from a more rapidly rotating superfluid component to the rest of the star \citep{ai75}.  
This component resides in regions of the interior where neutron vortices, which carry the angular momentum of the superfluid, are impeded in moving by pinning on crustal nuclei or on superconducting vortices in the core (or on both).
Since a superfluid in such conditions cannot slow down by outwards motion and expulsion of vortices, the superfluid component will retain a higher rotational frequency as the rest of the star slows down. 
A glitch occurs when vortices are suddenly unpinned and free to move outwards, allowing for a rapid exchange of angular momentum and the observed spin-up of the crust.

Catastrophic unpinning of vortices is expected once the velocity lag between the two components exceeds a maximum threshold, above which the pinning force can no longer sustain the hydrodynamic lift force exerted on the pinned vortices by the ambient superfluid. 
It has also been shown \citep{gl09, agh13} that, beyond some critical lag, a two-stream instability might develop and trigger the unpinning. 
 If in such events the lag is completely relaxed (or partially relaxed by a fixed amount) then the interglitch time interval corresponds to the time it takes for the system to reach the critical threshold again, driven by the nearly constant external torque. 
 Models relying on such a build-up and depletion of the superfluid angular momentum reservoir have been successfully used to explain the regular, similar glitches of some young pulsars \citep{accp93,pizz11, hps12}. 
 However this simple picture cannot account for the wide range of glitch sizes and waiting times between glitches seen in most pulsars.

 Glitch sizes in rotational frequency can range over four orders of magnitude in individual pulsars and appear to follow a power-law distribution \citep{mpw08}. 
This favours scale-invariant models of the dynamics of individual vortices in the presence of a pinning potential, such as the vortex avalanche model \citep{wm08} and the coherent noise model \citep{mw09}. 
 Alternative models involve non-superfluid mechanisms that can act as unpinning triggers before the critical lag is reached, such as crustquakes \citep{rud69,bp71} or heating episodes \citep{le96}. 
The crustquake-induced glitch model has been particularly favoured for the Crab pulsar as it may explain the persistent changes in slow-down rate observed after some of its glitches \citep{girp77,accp94} and could possibly lead to a power-law distribution of event sizes, similar to earthquakes.

A second type of irregularity is often seen in the rotational behaviour of pulsars, namely timing noise.
Thought to be partially caused by torque variations driven by two or more magnetospheric states \citep{lhk+10}, it manifests as a continuous and erratic wandering of the rotation rate around the predictions of a simple slow-down model.
While glitches are rapid and sporadic events in rotation rate, timing noise appears as a slow and continuous process.

Owing to observational limitations such as infrequent and irregular sampling and the presence of timing noise, the detection of glitches is an uncertain process. 
Moreover, the signature of timing noise in the data can be confused with glitches, so that the lower end of a glitch-size distribution is possibly contaminated by spurious detections.
Knowledge of this distribution is essential for any glitch theory.
The largest glitches are easily detected and can be used to constrain the minimum superfluid moment of inertia that can act as an angular momentum reservoir \citep{aghe12,chamel13}. 
The biases involved and the question of whether there is a minimum glitch size have not been addressed; so far, the smallest possible glitch has been assumed to lie below our detection limits.

In this paper we study the glitch detection capabilities of the current detection methods and define limits depending upon the intrinsic pulsar rotational stability, observing cadence and sensitivity. 
We apply these definitions to an extensive set of observations of the Crab pulsar and, by using an automated glitch detector, uncover the full glitch size distribution and show that there is a minimum glitch size.

\section{Limits on glitch detection}
To assess the level of completeness of the existing glitch samples, we quantify simple observational limits on glitch detection, applicable for a given pulsar and observing setup.
The first step towards this is establishing a working definition of what constitutes a glitch. 
Traditionally, glitches are identified by visual inspection of the pulsar's timing residuals, which are defined as the phase differences between measurements and the predictions of a model for the rotation.
To put this on a more formal footing, we define a glitch as an event characterised by a sudden, discrete positive change in rotational frequency ($\Delta\nu$) and a discrete negative or null change in frequency spin-down rate ($\Delta\dot{\nu}$).
These two sudden changes together make glitches distinguishable from timing noise \citep{elsk11,lhk+10}.

The timing residuals will be flat if the model describes the rotation of the pulsar well. 
For such a model, the timing residuals after a glitch at $t=t_g$ will follow a quadratic signature given by 
\begin{equation}
\label{gresids}
\phi_g=-\Delta\nu(t-t_g)-\Delta\dot{\nu}\frac{(t-t_g)^2}{2} \quad ; \quad(t>t_g)\,.
\end{equation}

\begin{figure}%[htbp]
\begin{center}
\includegraphics[width=84mm]{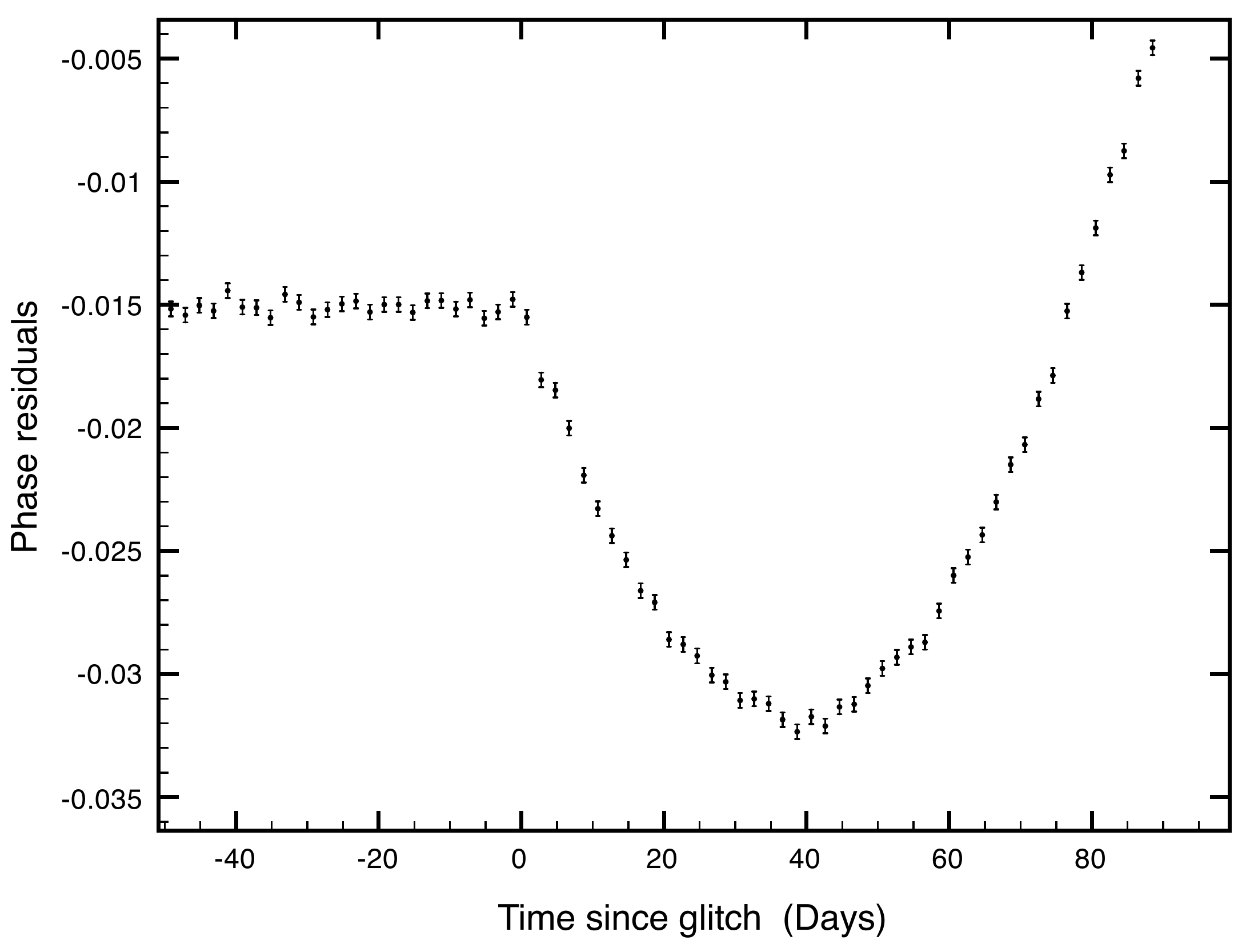} 
\caption{Example of a glitch signature in the timing residuals. 
The residuals are with respect to a model which describes well the rotation before the glitch ($t<0$ on the plot). 
This is a simulated glitch in the Crab pulsar's rotation, with $\Delta\nu=0.01$\,$\mu$Hz and $\Delta\dot{\nu}=-3.0\times 10^{-15}$\,Hz\,s$^{-1}$.}
\label{figS1}
\end{center}
\end{figure}

The frequency change $\Delta\nu>0$ produces a linear drift of the post-glitch residuals towards negative values, with the slope being the magnitude of the frequency step.
The effect of a change $\Delta\dot{\nu}<0$ is a parabolic signature which lifts the residuals towards positive values.
Therefore a glitch with a large, negative change in spin-down rate will produce positive residuals rising quadratically soon after the glitch (Fig.~\ref{figS1}).

Based upon these facts, we can define simple limits that describe our ability to detect glitches in the timing residuals.
%For a given $\Delta\dot{\nu}<0$ and for $\Delta\nu>0$ the residuals after the glitch are like those in Fig.~\ref{figS1}. 
If the observing cadence is not very frequent, it is possible that no observations occur before the post-glitch residuals rise above the extrapolation of the line defined by the pre-glitch ones (Fig. \ref{figS1}). 
This effect will primarily mask glitches with small $\Delta\nu$ and large $|\Delta\dot{\nu}|$. 
Requiring at least one observation before the rise of the residuals defines a minimum $\Delta\nu$ that can be detected (Eq. \ref{limits}).
If $\Delta\nu$ became smaller, the dip would become shallower and in the case that it is  undetectably small, the event is unlikely to be recognised as a glitch and might appear as timing noise.
To ensure detection, the maximum negative departure of the residuals ought to be larger than both the root mean square (RMS) of the timing residuals prior to the glitch and the typical error of the TOAs.
Therefore, a detectable glitch is a rapid event in which the effects of $\Delta\nu$ are recognisable over the effects of $\Delta\dot{\nu}$ and the limiting detectable value of $\Delta\nu$ depends on the observation cadence (one observation every $\Delta T$ days) and the largest of either the sensitivity of the observations or the typical dispersion of the timing residuals in rotational phase, $\sigma_\phi$, as
\begin{equation}
\label{limits}
\Delta\nu_\textrm{\scriptsize \,lim}=\rm{max} \left\{ \begin{array}{l}
	\Delta T|\Delta\dot{\nu}|/2 \\
	\\
   	\sqrt{2\,\sigma_\phi|\Delta\dot{\nu}|} 
	\end{array} \right . \quad \quad (\textrm{for}\quad \Delta\dot{\nu}<0).
\end{equation}

For simplicity and because of our particular focus on small events, any exponential recovery of the frequency, often observed after glitches, is not considered here. 
Nonetheless, we constructed several detectability curves, with decaying components and timescales similar to those observed in the Crab pulsar, and verified that our conclusions are not altered if exponential recoveries are present. 

These limits are consistent with the glitch samples of several pulsars, hence we believe they offer a realistic way to assess glitch detectability as it is commonly carried out.

\section{The Crab pulsar glitches}
\label{Crab}
The Crab pulsar (PSR B0531+21; PSR J0534+2200) is the central source of the Crab Nebula and a young neutron star widely studied since it was first observed in 1968.
The rotation of the Crab pulsar has been monitored almost every day for the last 29 years with the 42-ft radio telescope operating at $610$\,MHz at the Jodrell Bank Observatory (JBO) in the UK \citep{lps88,lps93}.
This offers an ideal dataset to test the completeness of the glitch sample because of its rapid cadence, good sensitivity and low dispersion of the timing residuals.

\subsection{Observations}
The product of each observation was the time of arrival (TOA) of one pulse at the observatory, corrected to the solar system barycenter.
The dataset comprises $8862$ TOAs starting in January 1984.
There is one TOA per day in general and two TOAs per day during some periods of time.
In addition, towards the beginning of the dataset, there are some isolated cases in which groups of TOAs are separated by up to $5$ days.
Finally, there are also a few gaps with no observations, generally no larger than $\sim20$\,days, when the telescope or observing hardware were unavailable due to maintenance.

The TOAs generally have errors of less than $0.001$ rotation, with more than $75$\% having uncertainties less than $0.0004$ rotation.
For groups of $20$ TOAs, which cover $20$ days on average, the timing residuals with respect to a simple slow-down model with two frequency derivatives typically give a dispersion similar to the TOA uncertainties (hence $\sigma_\phi\sim 0.0004$ rotations).

\subsection{Detection limits and the sample of detected glitches}
To study the glitch size distribution of the Crab pulsar we need a complete list of glitches for the time interval defined by the 42-ft dataset, and their main parameters $(\Delta\nu,\Delta\dot{\nu})$. As described above, we classify events as glitches based on the assumption that a glitch is a sudden, unresolved change in spin frequency, implying clearly defined features in the timing residuals.

We use the events included in the JBO online glitch catalogue\footnote{\url{http://www.jb.man.ac.uk/pulsar/glitches.html}}, which correspond to all the events published by \cite{elsk11} plus one new glitch that occurred on MJD\,$55875.5$ \citep{ejb+11}.
The event on MJD\,$\sim50489$, originally reported by \cite{wbl01}, was rejected because of its anomalous characteristics, already described by them.
No other glitches have been reported for this time-span by other authors \citep[e.g.][]{wbl01,wwty12} and we confirmed this by visually inspecting the timing residuals for all our dataset.
Our final list contains 20 glitches, with parameters covering the ranges 
$0.05\leq\left(\Delta\nu/\mu\textrm{Hz}\right)\leq6.37$ and 
$45\leq\left(|\Delta\dot{\nu}|/10^{-15}\textrm{Hz\,s}^{-1} \right)\leq2302$.
Here we use the glitch sizes reported by \citet{elsk11,ejb+11}.

We note the clear presence of four other glitches prior to the start of the 42-ft observations. 
However, the available data for that period is highly inhomogeneous and contains large gaps with no observations, making it difficult to define single detectability limits and complicating the use of the glitch detector (see below).
Hence, in order to work with a set of glitches that we know is statistically complete, we have not included them in our sample.  

Using $\Delta T=1$\,day and $\sigma_\phi=0.0004$ in Eq. \ref{limits} we find that all glitches (including the four early ones) show a clear separation from the detectability limits (Fig.~\ref{fig1}).
Thus, at least for intermediate and large glitches, we are uncovering the true $\Delta\nu$--$\Delta\dot{\nu}$ distribution, with no biases.

\section{The glitch detector}
\label{detector}
To confirm that we have identified every glitch in the data, especially small ones which may be missed by standard techniques, we developed an automated glitch detector to find and measure every timing signature that might be regarded as a glitch.
The detector assumes that a glitch occurred after every observation and attempts to measure its size, producing an output of glitch candidates (GCs) whenever $\Delta\nu>0$ and $\Delta\dot{\nu}\leq0$ are detected.

\subsection{Method}
\label{method}
The detector's technique is based on the fact that timing residuals, in the presence of timing noise or glitches, quickly depart from the best fit model of previous data, resulting in deviations from a mean of zero as newer observations are included and the model is not updated.
Below we describe the method step by step, optimised for the JBO dataset for the Crab pulsar, described above.
Different parameters should be used for different datasets. 

A fit for $\nu$, $\dot{\nu}$ and $\ddot{\nu}$ is performed over a set of $20$ TOAs using the timing software {\sc psrtime} and {\sc tempo2} \citep{hem06}, following standard techniques.
To test for a glitch occurring after the last TOA in a set, the timing residuals of the following $10$ TOAs, relative to that model, are fitted with a quadratic  function of the form of Eq. (\ref{gresids}) and separately with just the linear term in that equation.
The latter is to test the case $\Delta\dot{\nu}=0$.
The fit with the smallest reduced $\chi^2$ is selected.
When the quadratic fit is selected, an event is characterised as a GC only if the reduced $\chi^2$ is less than $15$, the quadratic part of the fit is negative ($\Delta\dot{\nu}<0$) and if the minimum of the fitted curve is at least $2.5$ times the dispersion of the timing residuals of the $20$ TOAs below zero. 
This last condition ensures a positive $\Delta\nu$ and a solid detection of its magnitude.
If the linear fit is selected, a new GC is created only if the reduced $\chi^2$ is less than $15$ and the slope of the fit is negative, indicating $\Delta\nu>0$. In this case the GC has a null or undetectable $\Delta\dot{\nu}$.
The conclusions of our analysis are not dependent on the choice of the maximum allowed reduced $\chi^2$ threshold\footnote{Changing this threshold to 20, for example, we obtained 12 new GCs, homogeneously distributed across the frequency range of GCs. There are no effects on the statistical results described in later sections.}.  The chosen value of $15$ is high enough to avoid missing signatures that one might regard as a glitch.  

The next step is to move the analysis forward by one TOA to define a new set of $20$ TOAs and test for a glitch occurring after this new TOA.
By doing this over the whole dataset, the dataset is explored for glitches after every single observation (with the exception of the first $19$ TOAs and the last 10 TOAs).

This method, however, causes some events to be detected multiple times.
This happens because the effects of $\Delta\nu$ and $\Delta\dot{\nu}$ may be detectable not only in the set of TOAs starting immediately after the event but also in some of the neighbouring trials.
Close inspection of the results shows that detections typically cluster in groups of $2$--$5$ trials, separated by no more than 2 days, and that clusters are typically $20$ to $30$ days apart.
To remove the repeated detections and produce a final list of GCs we select from each cluster of candidates the detection with the largest $\Delta\nu$ value. 
This is a conservative choice which makes the final list of GCs a representation of the maximum possible activity present in the data.
Also, this choice follows the experience gained from the detection of previously known glitches (section \ref{output}).

\section{Results}
We ran the detector over the 42-ft dataset, using the data from January 1984 to February 2013. 
The detector found all but one of the known glitches in this time-span as well as a large number of GCs.

\subsection {The output of the glitch detector}
\label{output}
\begin{figure}
\includegraphics[width=8.5cm]{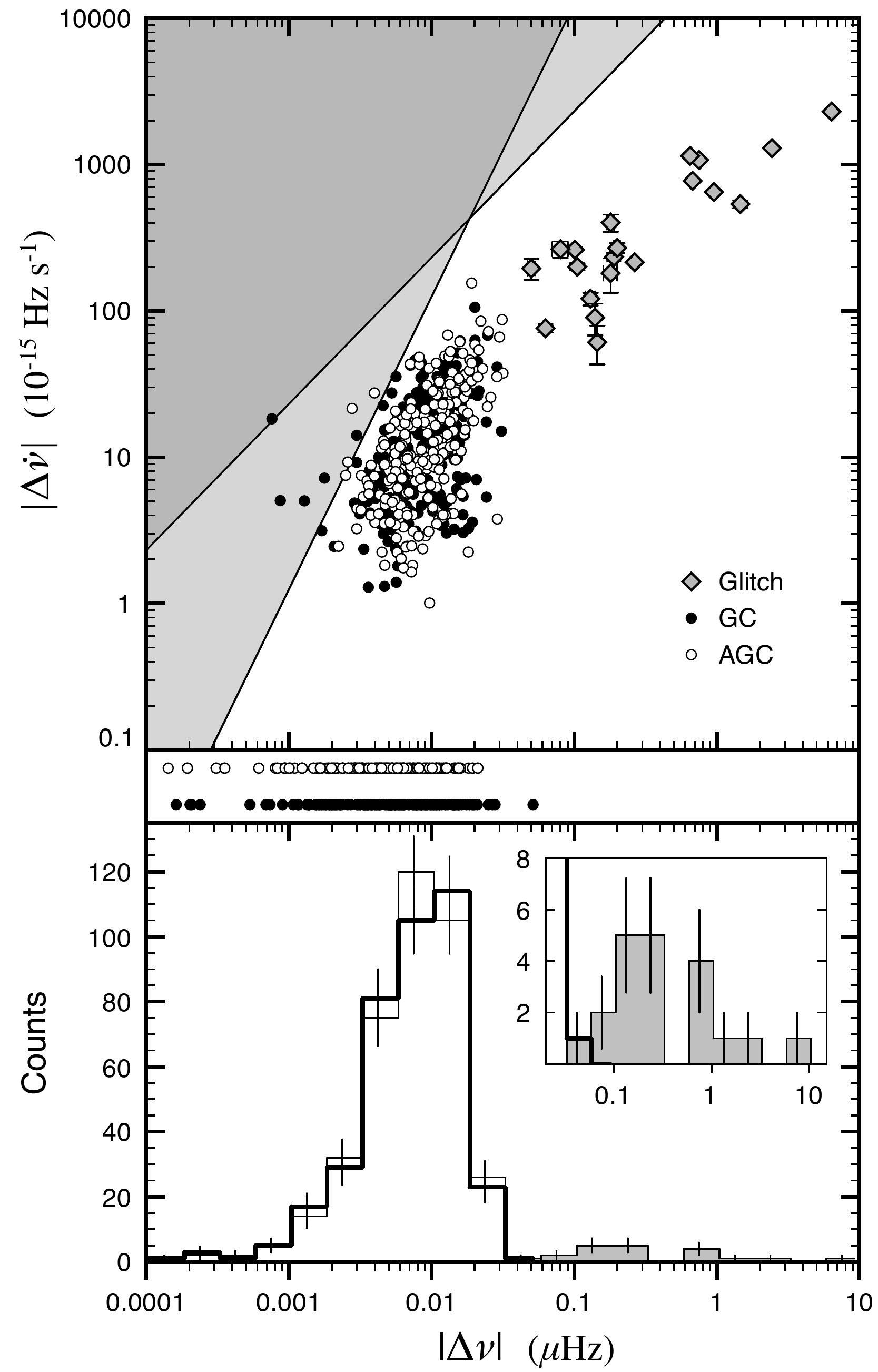} 
\caption{Previously known glitches (diamonds, from \citet{elsk11,ejb+11}), glitch candidates (GCs) and anti-glitch candidates (AGCs).
The top panel shows their distribution in the $|\Delta\nu|$--$|\Delta\dot{\nu}|$ plane. 
The straight line with the smallest slope represents the detection limit expected (Eq.~\ref{limits}) for a cadence of $\Delta T=1$\,day.
The one with the largest slope represents the detection limit expected for residuals of $\sigma_\phi=0.0004$.
Our observations are not sensitive to glitches in the shaded areas above these lines.
The middle panel shows the $|\Delta\nu|$ values of those candidates with undetectable $|\Delta\dot{\nu}|$.
The lower panel shows histograms for the $|\Delta\nu|$ values of the known glitches (filled grey), GCs (thick black) and AGCs (thin black). 
The inset shows a zoom in the region $|\Delta\nu|>0.02\,\mu$Hz.}
\label{fig1}
\end{figure}
\begin{figure*}
\includegraphics[width=17.7cm]{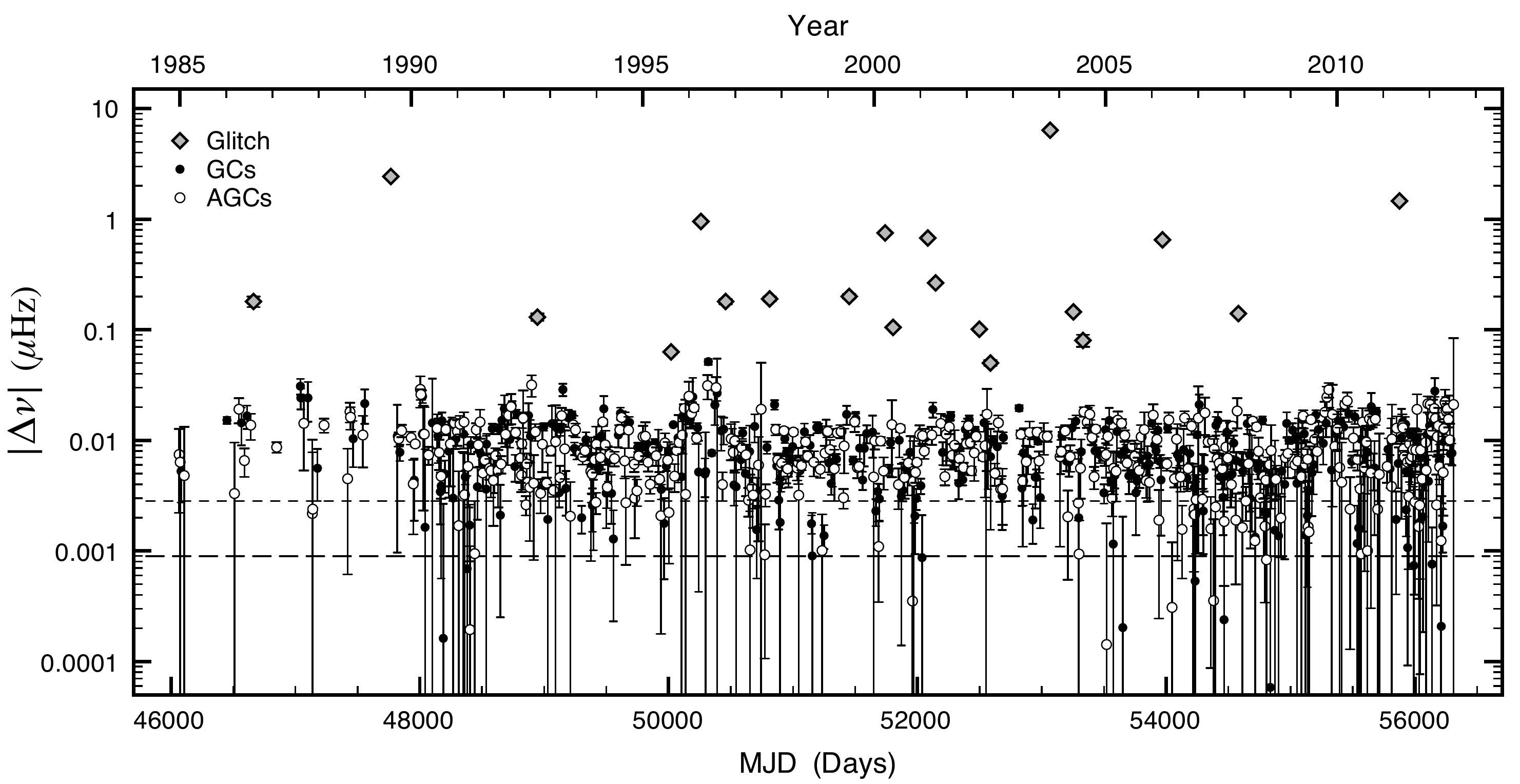} 
\centering
\caption{Time sequence of glitches, glitch candidates (GCs) and anti-glitch candidates (AGCs). 
Horizontal lines indicate the detection limits (Eq. \ref{limits}) for $|\Delta\dot{\nu}|=1\times10^{-15}$\,Hz\,s$^{-1}$ (long dashed) and $10\times10^{-15}$\,Hz\,s$^{-1}$ (short dashed). 
The low number of detections between the years 1985 and 1990 is caused by the presence of short gaps with no observations. }
\label{fig2}
\end{figure*}

%In order to simplify the results and to better control the computing times, the dataset was divided into several sections.
%The two largest known glitches were left in between sections, hence outside of the searched blocks of data. 
%We did not attempt to find the two largest glitches with the detector.
%However, $18$ glitches were included in the searched data, of which the detector found $17$. 
The only previously known glitch that was not found by the detector occurred on MJD\,$\sim52146.8$, only $63$ days after the previous glitch.
It was not labelled as a possible glitch because none of the fits, neither the quadratic nor the linear, gave a reduced $\chi^2$ less than $15$, one of the conditions to create a GC.
The smallest reduced $\chi^2$ among the fits around this glitch was $18$.
We attribute these poor fits to the influence of the recovery from the previous glitch.

The sizes $\Delta\nu$ and $\Delta\dot{\nu}$ that the detector measured for the known glitches are in good agreement with the values published by \cite{elsk11}.
Nevertheless, some differences can be found among the $\Delta\nu$ measurements.
As discussed above, because of the way the detector works, the effects of every glitch were detected in more than one set of TOAs.
The $\Delta\nu$ value coming from the set of TOAs offering the best fit (smallest reduced $\chi^2$) is always smaller than the published value, which is obtained by standard timing techniques.
In addition, this set of TOAs is normally the one starting one to three TOAs after the glitch epoch.
On the other hand, the glitch sizes obtained when testing at the correct glitch epoch are typically the largest and the most similar to the published values, though the fits have larger reduced $\chi^2$ values.
These effects are likely caused by unmodelled rapid exponential recoveries and were taken into account when selecting one candidate from a group of several candidates in the overall search.
The uncertainties of the GC sizes are the square root of the variances of the parameters, given by the Levenberg-Marquardt algorithm used to fit the data, multiplied by the square root of the reduced $\chi^2$ of the fit.

We reviewed the output of the glitch detector with the aim of producing a clean list of GCs. 
First, we removed from the original list all those GCs related to known glitches.
Then we kept only one candidate (the one having the largest $\Delta\nu$) per event, as mentioned in the description of the method (section \ref{method}).
Next, we visually inspected the timing residuals for all GCs having $\Delta\nu\geq0.02$\,$\mu$Hz and eliminated three which involved large data gaps or with timing residuals clearly contaminated by glitch recoveries.
We also examined the possibility that some GCs in this $\Delta\nu$ range could be caused by rapid changes in the electron density towards the Crab pulsar \citep{lps93}, which strongly affects the travel time of the pulsar emission at these low frequencies, introducing signatures in the data which can mimic a glitch.
To do so, we used observations taken at higher frequencies (mostly at $1400$\,MHz, with the Lovell telescope) and removed a further three GCs that were clearly caused by this effect.
However, the cadence of the Lovell observations is not as rapid as that of the 42-ft observations and we were unable to confirm some other possible cases of such non-achromatic events. 
We inspected the timing residuals of the largest remaining GCs ($|\Delta\nu|\geq0.02\,\mu$Hz) and found their signatures to be indistinguishable from timing noise,
though we acknowledge that discrimination between small glitches and timing noise is difficult. 
Nonetheless, in many cases no sharp transitions, typical of the known glitches, are observed at the GC epochs and the residuals are consistent with a smooth connection with the pre-GC-epoch residuals.

Our final list contains $381$ GCs. 
They are homogeneously distributed over the entire time-span and are clustered as a population in $\Delta\nu$--$|\Delta\dot{\nu}|$ space (Figs. \ref{fig1}, \ref{fig2}). 
The vast majority of them exhibit $\Delta\nu$ steps that are smaller than all previously detected glitches, leaving a gap between the $\Delta\nu$ distributions of real glitches and GCs which would be hard to populate with undetected events.

\subsection{Search for anti-glitches}
Given the distinct properties of the GC population, it is possible that the glitch-like signatures found by the detector are a component of the Crab pulsar's timing noise.
To test this idea and explore the noise nature of these irregularities, we performed a search for events with the opposite signature to a glitch, i.e. anti-glitch candidates (AGCs) with $\Delta\nu<0$ and $\Delta\dot{\nu}\geq0$, which are subject to the equivalent detection constraints as the normal glitches.
After removing repeated detections and $10$ events caused by glitch recoveries, gaps with no data and non-achromatic events (see above), we obtain $383$ AGCs. 
They are also separate from the glitch population and show very similar characteristics to the GCs (Figs. \ref{fig1}, \ref{fig2}).

\section{Discussion}
%\label{results}
\subsection{GCs and AGCs: glitches or timing noise?}
Using the Kolmogorov--Smirnov (K--S) test we can compare the $|\Delta\nu|$ distributions for GCs and AGCs, which are found to be statistically consistent with coming from the same parental distribution (with a K--S statistic of $D=0.037$ and $p_{KS}(D)=0.96$, thus a probability of only $\sim4\%$ for a false null hypothesis). 
The $|\Delta\nu|$ distributions for GCs and AGCs can be well described by  lognormal distributions, with probability density function (PDF) of the form
\begin{equation}
\label{LG}
p(|\Delta\nu|)=\frac{1}{\sqrt{2\pi}\sigma(|\Delta\nu|-\theta)}\exp{\left[-\left(\frac{\ln(|\Delta\nu|-\theta)-\mu}{4\sigma}\right)^2\right]} 
\end{equation}
for $|\Delta\nu|>\theta$, which gives $p_{KS}=0.85$ and $0.92$ respectively.
However, this result is only indicative since the lower ends of these distributions are not well probed by the observations (Figs. \ref{fig1}, \ref{fig2}). 

We also compared the $|\Delta\nu|$--$|\Delta\dot{\nu}|$ distributions of GCs and AGCs  using a 2-dimensional K--S test \citep{ptvf92}.
The test gives $D_\mathrm{2D}=0.084$, implying a probability of $\sim40\%$ that they come from the same distribution.
This relatively low probability is likely to be produced by differences in the $|\Delta\dot{\nu}|$ distributions between GCs and AGCs, since a K--S test over these two gives $p_{KS}=0.55$, considerably smaller than the one for $|\Delta\nu|$. 

Neither a power-law nor a lognormal distribution can describe well the joint $\Delta\nu$ distribution of the $20$ glitches plus all the GCs, with $p_{KS}<10^{-4}$. A power-law with a lower cut-off at $\Delta\nu\sim0.01\,\mu$\,Hz,  to account for the incompleteness of the sample at small sizes, gave a similarly poor fit.

Although it is possible that some of the GCs correspond to real glitches, we interpret all the above results as confirmation that the GCs and AGCs are generated by a symmetric noise process and that no new glitches have been found.
This timing noise component produces a continuous departure from a simple slow-down trend with variations that can be characterised by changes of $|\Delta\nu|\leq 0.03\,\mu$Hz and $|\Delta \dot{\nu}|\leq 200\times 10^{-15}$\,Hz\,s$^{-1}$.

\subsection{The glitch size distribution}
Having established that the 20 glitches form the complete sample of glitches the Crab pulsar has had in the last 29 years, we can address their statistical properties. 

\begin{figure}
\centering
\includegraphics[width=8.4cm]{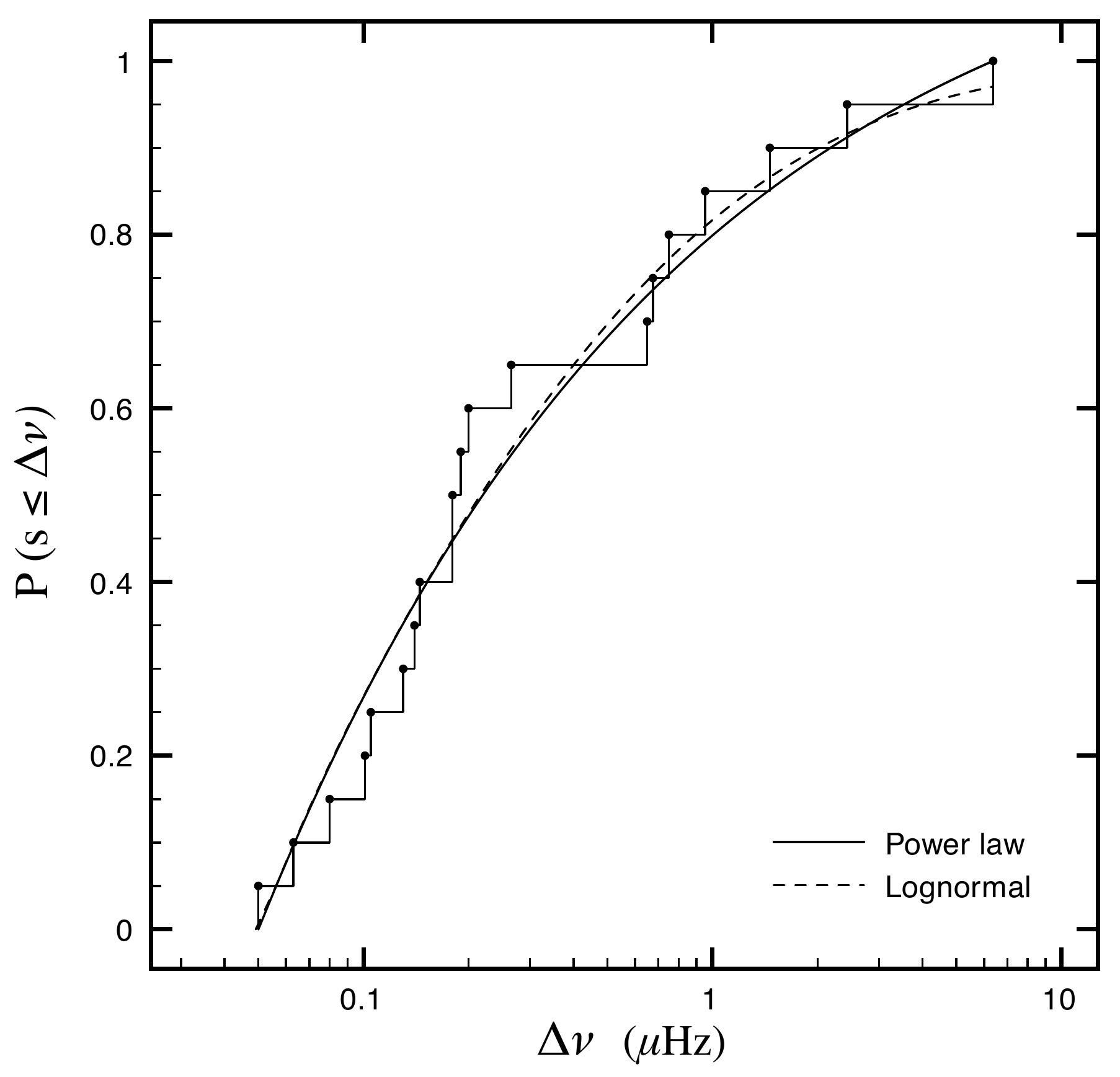} 
\caption{The cumulative distribution function of the observed glitch sizes, $\rm{s}$, and the corresponding power-law fit (solid line) given by Eq.~\ref{PL}, with $0.05\leq \Delta\nu\,(\mu$Hz$) \leq 6.37$ and $\alpha=1.36$. The dashed line corresponds to a lognormal fit (Eq.~\ref{LG}) with $\mu=-1.79$, $\sigma=1.9$ and $\theta=0.049\,\mu$Hz.}
\label{figS2}
\end{figure}

To determine the best-fit exponent $\alpha$ for a power-law PDF of the form
\begin{equation}
\label{PL}
p(\Delta\nu)=C{\Delta\nu}^{-\alpha}  \quad ,
\end{equation}
with $\Delta\nu_{\rm{min}}\leq\Delta\nu\leq\Delta\nu_{\rm{max}}$ and 
$C=(1-\alpha)(\Delta\nu_{\rm{max}}^{1-\alpha}-\Delta\nu_{\rm{min}}^{1-\alpha})^{-1}$, 
we use the maximum-likelihood estimator method. 
Setting $\Delta\nu_{\rm{min}}=0.05\,\mu$\,Hz and  $\Delta\nu_{\rm{max}}=6.37\,\mu$\,Hz, the values for the smallest and largest glitches observed respectively, we obtain $\alpha=1.36\,(+0.15,-0.14)$ (Fig.~\ref{figS2}). 
The value of the exponent does not depend strongly on the choice of limits, as long as these are a few times smaller or larger than the observed ones. 
To assess the goodness of the fit, we calculate the K--S statistic, $D=0.1$, and its probability value $p_{KS}(D)=0.9$, which corresponds to a $10\%$ probability that our null hypothesis (that the data follow the PDF described by Eq.~(\ref{PL})) is false. 
Thus our results confirm that the Crab glitch $\Delta\nu$ distribution is consistent with a power-law, a description motivated by theoretical models. 

If this power-law continued below $\Delta\nu_{\rm{min}}$, we would expect to have detected more than 10 glitches with $0.02\,\mu$Hz $<\Delta\nu<\Delta\nu_{\rm{min}}$ in the searched data, and the gap between glitches and GCs (in Figs. \ref{fig1} and \ref{fig2}) should have been populated. Thus we observe a rapid fall-off of the power-law for $\Delta\nu<\Delta\nu_{\rm{min}}$. 

However, the small sample size makes it impossible to exclude other distributions. 
For example, the same K--S probability is obtained for a lognormal distribution (Eq. \ref{LG}) with parameters $\mu=-1.79$, $\sigma=1.9$ and $\theta=0.049\,\mu$Hz, whose probability density function also quickly vanishes for $\Delta\nu<\Delta\nu_{\rm{min}}$. 
The same conclusions hold if the four glitches from before the start of this dataset are included in the sample.

Further confirmation of the rare occurrence of small glitches comes from the study of the  $|\Delta\nu|$--$|\Delta\dot{\nu}|$ distribution. 
Having shown that the latter is not affected by observational biases, the correlation between $|\Delta\nu|$ and $|\Delta\dot{\nu}|$ (apparent in Fig. \ref{fig1}) is confirmed to be a robust feature.
While $\Delta\nu$ measurements are very accurate, the acquired values of $\Delta\dot{\nu}$ are less certain and depend upon the method used to determine them, leading sometimes to large discrepancies. 
For this work we consistently calculated the glitch parameters for all 20 glitches, using the technique described in \cite{elsk11}. 
Using those measurements, the Spearman's rank correlation coefficient between $\Delta\nu$ and $|\Delta\dot{\nu}|$ is $\rm{rs}=0.776$ with $p(\rm{rs})=6\times10^{-5}$, which indicates a strong correlation. 
We note that the correlation becomes stronger if the four early glitches are included.
 
Given this relationship, any additional glitches would occupy a region of the $|\Delta\nu|$--$|\Delta\dot{\nu}|$ space well probed by the observations. 
Therefore, we observe a rapid decrease of the probability for glitch sizes below $\Delta\nu_{\rm{min}}$, which cannot be ascribed to incompleteness of the sample and hence indicates the existence of a minimum glitch size for the Crab pulsar.

\section{Implications for theoretical models}
%\label{discussion}
Such a limit for the smallest glitch size is challenging to our current understanding of glitches and has the potential to constrain the proposed mechanisms. 

Some simple considerations can be used to get a rough order of magnitude estimate for the number of neutron superfluid vortices that need to unpin to produce the smallest Crab glitch.
Each superfluid vortex carries a quantum of circulation $\kappa=h/2m_n\sim2\times10^{-3}$\,cm$^2$\,s$^{-1}$. 
Neglecting differential rotation of the superfluid (and entrainment), its total circulation at distance $r$ from the rotational axis will be $\Gamma=\oint v_s\cdot dl=N_v(r)\kappa=2\Omega_s(r) A$, where $N_v(r)$ is the number of vortices in the enclosed area $A$ and $\Omega_s$ is the superfluid angular velocity. 
Using $r=10^6$\,cm, the total number of vortices for the Crab pulsar is of the order $N_v\sim6\times10^{17}$. 
Conservation of the total angular momentum implies that the angular velocity change of the superfluid, $\delta\Omega_s$, relates to the observed glitch size $\Delta\nu$ by $\delta\Omega_s=2\pi\Delta\nu I_c/I_s$, where $I_c$ is the moment of inertia of the coupled component and $I_s$ is the superfluid moment of inertia that participates in the glitch. 
Using a typical value of $I_c/I_s\sim10^2$ and $\Delta\nu=\Delta\nu_{\rm{min}}=0.05\,\mu$Hz, the total number of vortices must be reduced by $\delta N_v\sim10^{11}$. 

The actual change in the superfluid angular momentum $L_s$ depends on the number of vortices that unpinned, the location and size of the region where this happened and the distance travelled by those vortices before they repin. 
For a more rigorous estimate, the change in $L_s$ can be approximated by $\Delta L_s\sim \hat{\rho}\kappa\xi R^3\delta N_v$, where $\hat\rho$ is the average density of the region involved, $R$ is the stellar radius and $\xi$ is the fraction of $R$ that unpinned vortices travel \citep{wm13}. For a typical value of $I_c\sim10^{45}$g\,cm$^2$ for the moment of inertia of the coupled component, the smallest glitch observed translates to an angular momentum change of $\Delta L_c=2\pi I_c \Delta\nu_{\rm{min}}=3\times10^{38}$\,g\,cm$^2$\,s$^{-1}$. 
Conservation of angular momentum leads to $\xi\delta N_v\simeq1.5\times10^9$  if one assumes typical values for the base of the crust, like $R=10$\,km and $\rho=10^{14}$\,g\,cm$^{-3}$.
Vortices are expected to repin after encountering a few available pinning sites, however as a conservative order of magnitude estimate we assume they cover a distance comparable to the thickness of the crust ($1.5$\,km) and take $\xi\leq0.15$, which means that at least $10^{10}$ vortices must unpin in a glitch with $\Delta\nu=\Delta\nu_{\rm{min}}$. 
Therefore the observed minimum glitch size, which is well above that expected for single-vortex unpinning events, implies the existence of a smaller characteristic length-scale which sets the lower cut-off for the range of the scale-invariant behaviour.  

The vortex avalanche model is based on the notion of self-organised criticality \citep[SOC,][]{btw87}, applications of which can be found for example in earthquake dynamics \citep{Her02} or superconducting flux-tube avalanches \citep{wwam06}. 
SOC occurs without the need of fine tuning of parameters, in several dynamical systems consisting of many interacting elements (the superfluid vortices in the case of a neutron star) which, under the act of an external slow driving force (the spin-down of the star), self-organise in a critical stationary state with no characteristic spatiotemporal scale. A small perturbation in such systems can trigger an avalanche of any size. Thus in the glitch avalanche model of \citet{wm08} vortex density is assumed to be greatly inhomogeneous and many metastable reservoirs of pinned vortices are formed, which relax independently giving rise to the observed spin-ups. Since such a system has no preferred scale the resulting glitch magnitudes follow a power-law distribution. This behaviour should however continue down to events involving the unpinning of only a few vortices, which is orders of magnitude below the observed cut-off.

The coherent noise model \citep{ns96} is a different, non critical mechanism which produces scale-free dynamics, even in the absence of interaction between the system's elements. 
In such systems a global stress is imposed to all elements coherently, to which they respond if it exceeds their individual unpinning threshold, giving rise to avalanches of various sizes. Both threshold levels (for each element) and stress strength are randomly chosen from respective probability distribution functions. 
The elements with thresholds smaller than the applied stress will participate in an avalanche and then be re-assigned new threshold values. 
New thresholds must always be assigned to a few elements, even when no avalanche is triggered, otherwise such a system will stagnate. 
A possible mechanism for this process in superfluids is the thermally activated unpinning of vortices \citep{mw09}, while the global Magnus force acts as the coherent stress.
The model predicts a minimum for the glitch magnitude, which represents the thermal creep only events, present even if all thresholds lie above the applied stress strength. But it also predicts an excess (with respect to the resulting power-law) of such small glitches, in contradiction to what we observe for the Crab pulsar. The lack of this overabundance of small events requires a broad distribution for the pinning potentials. \citet{mw09} studied the top-hat distribution and applied their model to the Crab pulsar. 
They found that the half-width of the distribution should be comparable to the mean pinning strength.
Even when such a broad distribution of pinning energies is introduced, independent unpinning of vortices as a random Poisson process of variable rate proves insufficient to produce scale-invariant glitches \citep{wm13}, indicating that the interaction between vortices and collective unpinning (a domino-like process) must be taken into account. The most prominent mechanism for collective unpinning is the proximity effect, in which a moving vortex triggers the unpinning of its neighbours. 
However such a mechanism requires extreme fine tuning, since power-law size distributions occur only if this effect is neither too weak (where thermal creep dominates) nor too strong (which always leads to large, system-spanning, avalanches) \citep{wm13}.

Another process which could lead to scale-invariant glitches are crustquakes \citep{Morl96}. 
Stresses develop in the solid crust of a neutron star because of the change in its equilibrium oblateness as the spin decreases, but also due to the interaction of the crustal lattice with the magnetic field and superfluid vortices in the interior. If the crust cannot readjust plastically it will do so abruptly when the breaking strain $\epsilon_{\rm{cr}}=\sigma_{\rm{cr}}/\mu$ is exceeded (where $\sigma_{\rm{cr}}$ is the critical stress and $\mu$ the mean modulus). 
This will result in both a spin-up (due to the moment of inertia decrease) and in a reaction of the superfluid \citep{accp96,rzc98}, which is evident in the post-glitch relaxation. The maximum fractional moment of inertia change associated with the $\Delta\nu_{\rm{min}}$ glitch is $|\Delta I|/I\leq 10^{-9}$; we note here that the glitch size can be significantly boosted by the crustquake induced unpinning of vortices \citep{ll99,Eich10}. 
Elastic stress on the crust due to change of the equilibrium oblateness builds up because of the almost-constant secular $\dot{\nu}$. 
Therefore the critical stress $\sigma_{\rm{cr}}$ will be reached in regular time intervals if all stress is relieved in each crustquake, and the total energy released will be $\Delta E_{\rm{el}}\propto\epsilon^2_{\rm{cr}}$. 
If the stress is only partially relaxed then the energy released will depend on the stress drop $\Delta\sigma$, and the time interval to the next crustquake will depend on the size of the preceding one. 
The latter correlation is observed for the glitches in PSR J0537-6910, which have been interpreted as crustquakes \citep{mmw+06}. For the Crab pulsar however, the lack of any such trends in our glitch sample indicates a more complicated picture.

\section{Conclusions}
\label{fin}
We have quantified our current glitch detection capabilities and, 
after a meticulous search for small glitches, we have shown that in the case of the Crab pulsar all glitches in this dataset have already been detected.
The full glitch size distribution exhibits an under-abundance of small glitches and implies a lower cut-off at $\Delta\nu\sim0.05\,\mu$Hz.
The existence of such a minimum glitch size implies a threshold-dominated process as their trigger, which still needs to be identified. 

Besides the occasional glitches, we have detected a continuous presence of timing noise having a well defined maximum amplitude, which can be described by step changes $|\Delta\nu|\leq 0.03\,\mu$Hz and $|\Delta \dot{\nu}|\leq 200\times 10^{-15}$\,Hz\,s$^{-1}$.  
The distinct properties of this noise component compared to the glitches imply that timing noise cannot be attributed solely to unresolved small glitches produced by the exact same mechanism.

\section*{Acknowledgments}
Pulsar research at JBCA is supported by a Consolidated Grant from the UK Science and Technology Facilities Council (STFC). 
C.M.E. acknowledges the support from STFC and FONDECYT (postdoctorado 3130512).
D.A. and A.L.W. acknowledge support from an NWO Vidi Grant (PI Watts).

\bibliographystyle{mn2e-mod}
\bibliography{journals,modrefs,psrrefs,crossrefs,new1}

\label{lastpage}
\end{document}